\documentclass[sigconf,screen,authorversion]{acmart}

\usepackage{hyperref}
\usepackage{subcaption}
\usepackage{booktabs}
\usepackage{multirow}
\usepackage{graphicx}
\usepackage[inline]{enumitem}

\AtBeginDocument{%
  \providecommand\BibTeX{{%
    \normalfont B\kern-0.5em{\scshape i\kern-0.25em b}\kern-0.8em\TeX}}}

\copyrightyear{2025} 
\acmYear{2025} 
\setcopyright{cc}
\setcctype{by}
\acmConference[ETRA '25]{2025 Symposium on Eye Tracking Research and Applications}{May 26--29, 2025}{Tokyo, Japan}
\acmBooktitle{2025 Symposium on Eye Tracking Research and Applications (ETRA '25), May 26--29, 2025, Tokyo, Japan}\acmDOI{10.1145/3715669.3723129}
\acmISBN{979-8-4007-1487-0/2025/05}

\begin{document}

\title[Eye Movements as Indicators of Deception]{Eye Movements as Indicators of Deception: A Machine Learning Approach}

\author{Valentin Foucher}
\authornote{Both authors contributed equally to the paper}
\affiliation{%
  \institution{General Psychology,
  Faculty of Engineering, Computer Science, and Psychology,
Ulm University}
  \city{Ulm}
  \country{Germany}
}
\email{valentin.foucher@uni-ulm.de}
\orcid{0009-0000-2632-3519}

\author{Santiago de Leon-Martinez}
\authornotemark[1]
\affiliation{%
  \institution{Faculty of Information Technology, Brno University of Technology}
  \city{Brno}
  \country{Czechia}
}
\additionalaffiliation{%
  \institution{Kempelen Institute of Intelligent Technologies}
  \city{Bratislava}
  \country{Slovakia}
}
\email{santiago.deleon@kinit.sk}
\orcid{0000-0002-2109-9420}

\author{Robert Moro}
\affiliation{%
  \institution{Kempelen Institute of Intelligent Technologies}
  \city{Bratislava}
  \country{Slovakia}
}
\email{robert.moro@kinit.sk}
\orcid{0000-0002-3052-8290}

\renewcommand{\shortauthors}{Foucher, et al.}

\begin{abstract}
Gaze may enhance the robustness of lie detectors but remains under-studied. This study evaluated the efficacy of AI models (using fixations, saccades, blinks, and pupil size) for detecting deception in Concealed Information Tests across two datasets. The first, collected with Eyelink 1000, contains gaze data from a computerized experiment where 87 participants revealed, concealed, or faked the value of a previously selected card. The second, collected with Pupil Neon, involved 36 participants performing a similar task but facing an experimenter. XGBoost achieved accuracies up to 74\% in a binary classification task (Revealing vs. Concealing) and 49\% in a more challenging three-classification task (Revealing vs. Concealing vs. Faking). Feature analysis identified saccade number, duration, amplitude, and maximum pupil size as the most important for deception prediction. These results demonstrate the feasibility of using gaze and AI to enhance lie detectors and encourage future research that may improve on this.
\end{abstract}




\maketitle

\section{Introduction}
In a world where the lines between truth and lies are increasingly indistinct, the development of reliable deception detection has become paramount. Traditional methods of detecting deception have relied on the measurement of physiological signals, such as heart rate, respiration rate or skin conductance while asking crime-relevant questions \cite{raskin_psychopathy_1978}. However, these techniques are not only intrusive but also demonstrate low reliability and can easily be manipulated by experienced deceivers \cite{ben-shakhar_critical_2002}. A category of test called Concealed Information Test (CIT), emerged in crime contexts to undermine concealed knowledge \cite{ben-shakhar_validity_2003}. This test aims to examine participants' reactions to the presentation of crucial information to detect unspoken information. Traditionally focused on response time, this test has been taking advantage of the pervasive accessibility and ease of use of eye-tracking devices to integrate eye movements with the objective of enhancing its resistance to countermeasures.

\subsection{Related Works}

Concealing a known face has been found to correlate with a decrease in the number of fixations on the concealed face along with an increase in fixation duration \cite{schwedes_revealing_2012, millen_tracking_2017}. Similarly, fewer but longer fixations were observed on crime-related details when viewing images related to a mock crime \cite{peth_fixations_2013}, and fewer fixations were directed toward an accomplice when deceiving during online interview \cite{pak_eye_2013}. Deception would also influence saccades, with a higher saccade frequency and velocity when lying \cite{vrij_saccadic_2015}, but smaller saccade amplitudes \cite{lim_lying_2013}. However, the direction of their effect remains ambiguous, and contradictory outcomes have been reported in two analogous questionnaire studies \cite{fang_assessing_2021}. Research on the impact of deception on blinking behaviours consistently shows a decrease in the number of blinks during interviews or when concealing the recognition of a target stimulus compared to control stimuli \cite{fukuda_eye_2001, leal_blinking_2008, leal_occurrence_2010, peth_fixations_2013}. The impact of deception on blink duration remains ambiguous, with some evidence suggesting a decrease in maximum blink duration when lying in response to mock-crime-related questions versus responding truthfully \cite{marchak_detecting_2013}. Pupil size has also been demonstrated to be a potential indicator of deception. Research has shown that greater pupil dilation occurs in response to false statements made in contexts involving benign scenarios \cite{dionisio_differentiation_2001} or when reading statements about a mock crime that participants committed \cite{webb_eye_2009, cook_lyin_2012}. Greater pupil dilation has also been observed in response to the presentation of mock-crime-related items \cite{lubow_pupillary_1996, twyman_systems_2013, proudfoot_deception_2015}, familiar faces \cite{seymour_combining_2013}, personal names \cite{chen_concealed_2023}, or selected numbers \cite{foucher_unveiling_2024} despite concealing them. 

Several computational models have demonstrated the predictive contribution of facial and eye parameters in detecting lies from truths (\cite{prome_deception_2024} for a recent review) as evidenced by their ability to discriminate authentic from deceptive online reviews \cite{ott_finding_2011, lai_human_2019} or truthful from deceptive statements in real-life trials \cite{perez-rosas_deception_2015}. Indeed, automated deception detection systems exposed the role of non-verbal behaviours, including eye movements, during interviews \cite{oshea_intelligent_2018}. Eye movements appeared to be even more effective at predicting deception than facial micro-movements with an accuracy of 78\% \cite{khan_deception_2021}, reporting blinks as the strongest indicator of deception and confirming the results of a previous study that developed a new blink rate metric to differentiate lies from truths \cite{borza_eye_2018}. Saccades however were not found of importance in their model. Fixations and pupil size could also predict deception in questionnaire surveys with an accuracy of 74\% \cite{fang_assessing_2021}. 

\subsection{Present Study}

These models sought to differentiate lies from truths. However, to our knowledge, no predictive models have attempted to detect the concealment of information, which would be particularly advantageous in the context of the CIT. Furthermore, concealing or faking information has been shown to induce different eye movement behaviours \cite{foucher_unveiling_2024} and may be identified differently in deception recognition systems. The present study aims to fill this gap by evaluating the efficacy of various eye parameters in predicting deception in CIT across two datasets. In both datasets, participants were instructed to select a card value between 1 and 6, and were then presented with all the possible card values on a screen. Participants were instructed to employ only their eyes to either reveal, conceal, or fake the value of their card to their interlocutor. The primary objective of this study was to ascertain the extent to which eye parameters can predict deceptive intentions. The second objective was to determine which eye parameters are the best suited for such prediction. A feature analysis was performed to assess the importance of each feature in the prediction of the deceptive intentions.

\section{Experimental Methods}
The prediction of deceptive intentions was tested in two different contexts. The Eyelink dataset (collected by the Eyelink 1000) comprises an experiment involving participants deceiving in front of a computer with a chinrest. The Neon dataset (collected by the Pupil Neon) corresponds to an experiment involving participants deceiving an experimenter. The key differences between the two contexts are: 1) the target of the deception (who the participants are lying to) is either a computer or a physical person and 2) the Pupil Neon is a wearable eye tracker that allows much more freedom of movement of the head as compared to a chinrest and was used to improve the ecological validity of the design.

\subsection{Design}

\begin{figure*}[ht]
  \centering
  \includegraphics[width=0.85\linewidth]{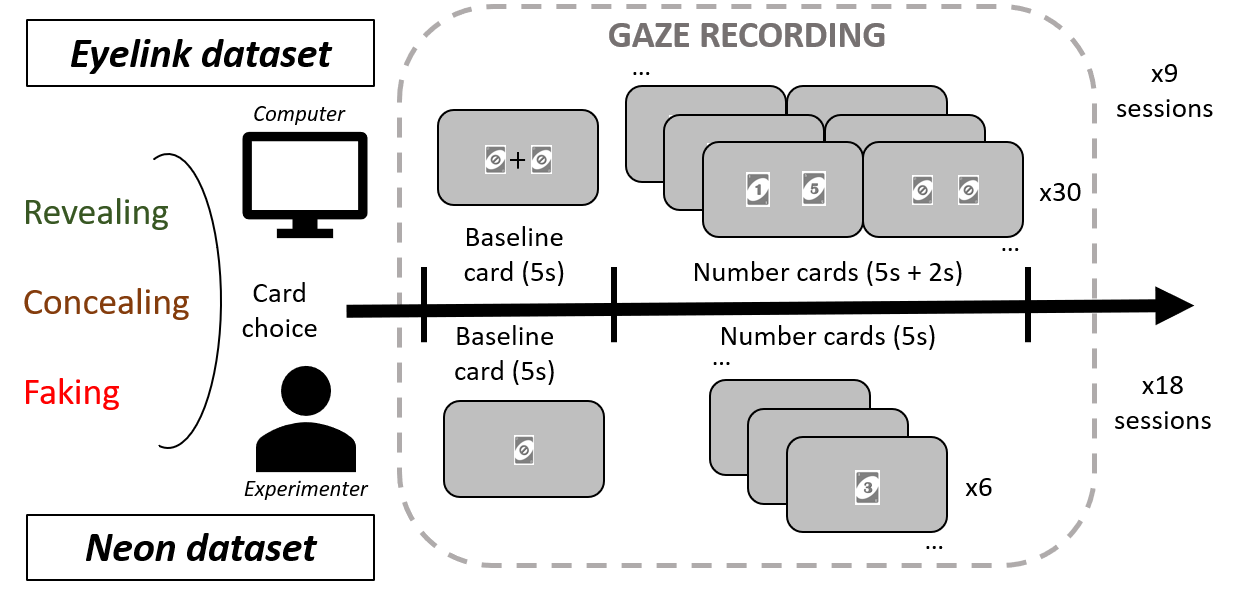}
  \caption{Experiments from both datasets contain the selection of a target, followed by the presentation of a baseline card and the number cards one after another for 5s each. In the Eyelink dataset, the number cards are separated by baseline cards for 2s. Participants are instructed to use their eyes to reveal, conceal, or fake the value of their target to either a computer or an experimenter.}
  \label{fig:design}
  \Description{Experiments from both datasets contain the selection of a target, followed by the presentation of a baseline card and the number cards one after another for 5s each. In the Eyelink dataset, the number cards are separated by baseline cards for 2s. Participants are instructed to use their eyes to reveal, conceal, or fake the value of their target to either a computer or an experimenter.}
\end{figure*}

\paragraph{Eyelink Dataset.} Eyelink dataset comprises an experiment in which participants were instructed to play a homemade card game with an interlocutor who could be a partner, an opponent, or a referee, represented on a computer by a partner smiley, an opponent smiley and a referee smiley respectively. In this game, participants began a session with six blue UNO cards, numbered from 1 to 6, placed randomly side by side in front of them. After selecting a number -- called the target -- participants were informed about the interlocutor with whom they would be interacting. Pairs of numerical cards were then displayed on the screen. The participants’ objective was to interact with the interlocutor regarding the target using only their eyes, with the interlocutors described as algorithm-driven systems designed to read the participants' eye movements. There were three types of instructions:
\begin{enumerate*}
\item  \textit{Revealing the truth} is the interaction with the partner: participants have to disclose the target to them. For instance, if they selected the number ‘2’, they want the partner interlocutor to realise that they selected this number.
\item  \textit{Faking the truth} is the interaction with the opponent: participants have to mislead them into guessing another card than the target. For instance, if they selected the number ‘2’, they want the opponent interlocutor to guess another number, like the number ‘5’. The choice of this fake number was up to the participant.
\item  \textit{Concealing the truth} is the interaction with the referee: participants have to prevent them from guessing the target. For instance, if they selected the number ‘2’, they do not want the referee interlocutor to realise that they selected this number.
\end{enumerate*}

Thirty pairs of numbers were presented 5s each with pseudo-randomized positions and orders. Each number appeared ten times. A fixation cross accompanied by a pair of neutral cards was presented between each pair of numbers for 3s to allow the pupil size to return to a baseline level. Nine sessions comprised interactions with each of the three randomly ordered interlocutors, leading to the presentation of 270 pairs in total. Each session started with an eye-tracker calibration during which participants were instructed to follow a dot’s movement from the centre of the screen to six predetermined positions along the borders. The requirement to communicate exclusively through eye movements was implemented to minimise body movements during the experiment, while ensuring participants maintained eye contact with the screen. No guidelines were provided concerning the means to achieve the desired behaviour leaving the participants to develop their own strategies for conveying the intended information to their interlocutor.

\paragraph{Neon Dataset.} The Neon dataset constitutes an experiment that is analogous to the one presented in the Eyelink dataset, with the exception that participants engaged in a game against an experimenter seated in front of them, who poses as a mentalist as opposed to a computer. A screen was positioned centrally on the table to present visual stimuli, and a computer was situated adjacent to the participants to collect their responses. At the beginning of each round, participants were presented with six UNO cards numbered from 1 to 6 on the screen and were required to select one as the target. Thereafter, a baseline card was displayed for 5 seconds, followed by the presentation of all six number cards in a random order, one after another, at the centre of the screen for 5 seconds each. As in the preceding experiments, the objective for the participants was to interact with the experimenter regarding the value of their target just by using their eyes. The three interactions from the previous experiments could occur, also referred to as “Revealing”, “Faking”, and “Concealing” the truth, except that the interlocutor was the experimenter who behaved as either a partner, an opponent or a referee. The experimenter was unaware of the value of the target and the interaction participants were performing. Participants were instructed to communicate with the experimenter according to their designated interaction’s instruction employing only eye movements, without any requirement to look at any specific location or object. Eighteen sessions comprised the three balanced interaction instructions, leading to the presentation of 108 cards in total. It is noteworthy that no calibration was necessary to record data with the Neon system. The experimental design from both datasets is illustrated in Fig.~\ref{fig:design}.

\subsection{Apparatus \& Stimuli}
In the Eyelink dataset, eye movements were measured using the EyeLink 1000 from SR Research Ltd with a 1000Hz sample frequency. The setup consisted of one table supporting a screen and the eye-tracker. Participants were seated at a distance of 57cm from a screen in a room with low ambient light, with their heads on a chin rest to prevent noise movements. The instructions were white-coloured and presented against a grey background on a display monitor (54”, 1920x1080px, 100Hz). The interaction instructions given to participants regarding their interlocutor's role were conveyed through the utilisation of emoticons (e.g. a green smiley face indicated the partner, an orange angry face indicated the opponent and a yellow neutral face indicated the referee). The use of emoticons in the instructions was not part of the eye-tracking recordings. 

In the Neon dataset, eye movements were recorded using the Neon glasses from Pupil Labs GmbH with a 200Hz sample frequency and the “Just act natural” frame. Neon glasses have the specific property of not requiring calibration to record gaze data. The participant and the experimenter faced each other at a distance of 100cm, in a room with ambient lighting. Between them, a screen (31x17.5cm, 1920x1080px, 60Hz) was placed horizontally on the table at a distance of 50cm from both the participant and the experimenter, ensuring that the stimuli material for the experiment remained constantly visible.

In both datasets, UNO cards were used due to their unambiguous design which facilitated immediate identification of the number. The numbers 1 to 6 were used, with the Skip card serving as the baseline. The numeric UNO cards displayed on the screen were created by numeric replicas transposed in grey and white colour with simple pixel discrimination, and displayed on a grey background to minimize the differences in brightness distribution in the stimuli. In the Eyelink dataset, the area of interest associated with each card was defined with an additional degree of visual angle around the borders of the card, defining a rectangle of 7.9 x 6.1 degrees of visual angle. In the Neon dataset, the area of interest associated with each card was defined using Pupil Labs markers at the tablet’s corners.

\subsection{Participants}
\label{sec:methods:participants}
The experiments were approved by the Ethics Committee of Ulm University. Prior to participation, all participants were informed of the option to discontinue their involvement at any stage and provided written consent. For the Eyelink dataset, the study was publicised using the local university study management system %
and participants received credit hours as compensation for their participation. For the Neon dataset, the study was publicised through an external recruitment agency and participants received a 20€ voucher as compensation for their participation. The prerequisites for participation included in both cases normal or corrected-to-normal vision and the absence of epileptic seizures. The Eyelink dataset contains 87 participants, including 56 females (M=22.1yo, sd=2.1yo). The Neon dataset contains 36 participants, including 21 females (M=34.3 yo, sd=8.8).

\subsection{Preparation of the Data}
The datasets contain eye movement data that was exported from the Eyelink Data Viewer software package (Eyelink dataset) \cite{sr_research_ltd_eyelink_2022} and the Pupil Labs Cloud (Neon dataset), and were preprocessed on R \cite{r_core_team_r_2023}. Data is organised by participants and trials, with each trial corresponding to the presentation of a card. To capture meaningful changes, only trials containing a target were considered. To remove artefacts in the eye movements’ definition, usual thresholds were used: fixations shorter than 60ms were removed, as blinks shorter than 60ms and longer than 700ms, and saccades shorter than 15ms and longer than 400ms \cite{rayner_eye_1998, foucher_unveiling_2024}. Fixations longer than 5s were also removed since cards were presented for 5s each. Following recent pupillometry recommendations \cite{mathot_methods_2022}, pupil information was computed from the gaze located on the cards during the first 2.5s after the card presentation, including a cubic interpolation on the Eyelink data to account for missing data due to blinks. Baselines were defined by the 50ms before the presentation of each card, enabling the calculation of a baseline-corrected pupil signal. Outliers with an absolute Z score above 3 were excluded. Data was finally down-sampled from 1000Hz to 20Hz by averaging pupil size in 50ms-bins to capture meaningful size variations.

As the literature does not provide clear evidence that would rule out a specific eye movement, a comprehensive array of fundamental first-order features that characterize all these eye movements was tested. The resulting dataset for each trial and participant comprised the following features: fixation number and duration, saccade number, duration and amplitude, blink number and duration, mean, maximum and minimum pupil size, and the mean pupil size over 2500ms in 50ms-windows separated into fifty bins. Consequently, 60 eye parameters were investigated and could be separated into 7 eye-movement-related features and 53 pupil-related features.

\section{Prediction of Deceptive Intentions}

Since the experiments are Concealed Information Tests, originally designed to recognize the concealing of information, we perform the following predictive tasks: 1) predict the trials where the participants concealed the value of the target, compared to when they revealed it to the interlocutor, and 2) consider a third type of deceptive intention by predicting the trials where the participants concealed, faked or revealed the value of the target to the interlocutor. Therefore, task 1 is a binary classification between Concealing and Revealing the truth, while task 2 is 3-class classification between Concealing, Faking, and Revealing the truth. The XGBoost classifier was used for all predictive tasks.

\subsection{Datasets}

\paragraph{Target Distribution.} The Eyelink dataset contained the following instances of the following classes:  1996 Concealing, 2065 Faking, and 1996 Revealing. The Neon dataset contained the following instances of the following classes:  160 Concealing , 161 Faking, and 181 Revealing. Due to the reasonable balance of the data, in both the binary classification and 3-class classification no sample or loss weighting/balancing was used for the predictive models. 

\paragraph{Feature Datasets.} To allow a better understanding of which parameters benefit the performances of the model, and to account for the bigger proportion of the pupil size in the number of parameters, the features were separated into three groups: 1) one including only eye movements (fixations count \& duration, blink count \& duration, saccade count \& duration \& amplitude), 2) one including only pupil size (mean, maximum, minimum, fifty time-bins of 50ms), and 3) one including all the eye parameters. Both the Eyelink and Neon datasets were separated into the 3 separate feature datasets for resulting 6 datasets. Each of these were then used for both predictive tasks resulting in 12 conditions, each with 5 replications with different train/test splits. This resulted in 60 trained and tuned final models and a total of 150,600 trained models across all conditions and replications (including cross validation). All conditions and their replications follow the same methodology (splitting, hyperparameter tuning, and evaluation) within and across datasets. 

\subsection{Model and Training Methodology}

\paragraph{Dataset Splitting and Training.} Datasets were first split into 5-folds stratified by participant ID (all data of one participant was guaranteed to be in only one fold) of roughly 20\% each. These provide 5 different replications (per each of the 12 conditions) each using one of the 5 different test folds (20\%) and the remaining data for train and validation (80\%). Therefore, 5 different final tuned models were trained and evaluated on 5 different train/validation and test sets for one dataset, where one of these 5 we call a replication. 

After a train/validation and test set were fixed, the train/validation dataset was split (stratified by participant ID) into a train and parameter validation set (70\%) and early stopping (for \# of trees) validation set (10\%). Finally, the train and parameter validation set (70\%) was split into stratified 5-folds for cross validation of hyperparameters, resulting in a train set (56\%) and parameter validation set (14\%). Each combination of hyperparameters were trained and evaluated on the 5 different folds and averaged across the 5 to determine the best hyperparameter set. For efficiency, early stopping (for \# of trees) was also performed for each fold, but was evaluated on the same parameter validation set and only used to find the best score for that fold for a selection of parameters and later discarded (instead of being added to the best hyperparameter set).

Once the best hyperparameter set was found, a new model was trained on the train and parameter validation set (70\%) using those parameters. Then the early stopping validation set (10\%) was used to determine the best number of trees. With the best hyperparameter set and the best number of trees found from early stopping, a final model was trained on the whole train/validation set (80\%) and then evaluated on the left out test set (20\%). 

\paragraph{Evaluation Metrics.} Accuracy was used to determine the best hyperparameter set  by averaging accuracy scores of the 5 cross validation splits. Logloss (mlogloss in 3-class) was used for finding the best iteration in all cases of early stopping. For test evaluation, we report the average of the 5 replication accuracy scores achieved on the test sets along with standard deviation. 

\paragraph{Model: XGBoost Classifier.} The model used for all conditions was the XGBoost classifier \cite{10.1145/2939672.2939785}. XGBoost is a gradient boosting algorithm that builds ensembles of decision trees that predict the appropriate class. It is well known inside and outside of the machine learning community for its performance and efficiency on tabular data, compared to other approaches. Like other decision tree models it does not require standardization of features and provides methods for measuring the importance of features in predicting a target. Initially, we also used tabNet \cite{arik2020tabnetattentiveinterpretabletabular}, an explainable deep learning model for tabular data, but abandoned it due to poor results when compared to XGBoost. 

\paragraph{Hyperparameters.} All models were set to have a maximum of 10,000 trees (parameter: n estimators) with 35 early stopping rounds except for the final model, which used the best number of trees from the early stopping validation set and did not have early stopping. For hyperparameter selection, 100 combinations were sampled from the following hyperparameters and their distributions: learning rate (uniform from 0.001-0.191), subsample (uniform from 0.5-1.0), colsample by tree (uniform from 0.5-1.0), min child weight (1, 3, 5, 7), max depth (0, 3-15), alpha (0, 0.01, 1, 2, 5, 7, 10, 50, 100), lambda (0, 0.01, 1, 2, 5, 7, 10, 50, 100).

\subsection{Results of Classification and Feature Importance}

The results are shown in Table~\ref{tab:results}. As a baseline, we use a dummy model that  predicts the proportion of the majority of classes found in the whole dataset rather than reporting the proportion for each train and test replication that varies only slightly. After evaluation of all 60 final models (5 replications of each 12 conditions), a feature importance analysis for each model was performed. We use a popular game theoretic approach of Shapley (SHAP) values to determine the importance of each feature in explaining the prediction \cite{shapley:book1952}. For each data point in the test set, a SHAP value is calculated for each feature representing the contribution of the feature in prediction. To find the importance of a particular feature in predicting the output, the values are summarized by taking the absolute value and the mean for each feature. The values themselves cannot be compared across models (without normalization or other processing). In order to compare the 60 final models and determine shared important features, we take the top 5 nonzero features for each model and show their occurrence across the 5 replications for each of the 12 conditions.

\begin{table*}
\centering
\caption{Results for the 12 conditions, calculated by averaging the test accuracy of the final models for each of the 5 replications. Additionally, the top 5 nonzero features are shown for each of the 5 replications and their occurrence in parentheses. Note that the maximum occurrence possible is (5) meaning that every final model of all 5 replications found the feature to be important and we only show features that have (2) or more occurrences across the 5 replications.}
\label{tab:results}
\resizebox{\textwidth}{!}{%
\begin{tabular}{@{}l|l|ll|ll|ll@{}}
\toprule
\multirow{2}{*}{\textbf{Revealing / Concealing}} &
  Baseline &
  \multicolumn{2}{c|}{All Features (n=60)} &
  \multicolumn{2}{c|}{Eye-movement Features (n=7)} &
  \multicolumn{2}{c}{Pupil Features (n=53)} \\
 &
  \multicolumn{1}{c|}{\begin{tabular}[c]{@{}c@{}}Majority\\ Class\end{tabular}} &
  \multicolumn{1}{c}{Avg Acc (std)} &
  \multicolumn{1}{c|}{Top 5 Features (\textgreater{}1)} &
  \multicolumn{1}{c}{Avg Acc (std)} &
  \multicolumn{1}{c|}{Top 5 features (\textgreater{}1)} &
  \multicolumn{1}{c}{Avg Acc (std)} &
  \multicolumn{1}{c}{Top 5 Features (\textgreater{}1)} \\ \midrule
Eyelink dataset &
  0.500 &
  0.736 (0.015) &
  \begin{tabular}[c]{@{}l@{}}Saccade amplitude (5)\\ Saccade number (5)\\ Saccade duration (5)\\ Max pupil size (3)\\ Fixation duration (3)\end{tabular} &
  0.733 (0.026) &
  \begin{tabular}[c]{@{}l@{}}Saccade amplitude (5)\\ Saccade number (5)\\ Saccade duration (5)\\ Blink duration (3)\\ Fixation duration (3)\\ Fixation number (2)\\ Blink number (2)\end{tabular} &
  0.585 (0.024) &
  \begin{tabular}[c]{@{}l@{}}Min pupil size (5)\\ Pupil window 801 (5)\\ Max pupil size(4)\\ Pupil window 1001 (3)\\ Mean pupil size (2)\\ Pupil window 2301 (2)\end{tabular} \\ \midrule
Neon dataset &
  0.531 &
  0.638 (0.069) &
  \begin{tabular}[c]{@{}l@{}}Saccade amplitude (5)\\ Max pupil size (5)\\ Saccade duration (3)\\ Pupil window 951 (2)\\ Pupil window 1451 (2)\end{tabular} &
  0.581 (0.033) &
  \begin{tabular}[c]{@{}l@{}}Saccade amplitude (5)\\ Saccade duration (5)\\ Fixation duration (4)\\ Saccade number (3)\\ Blink duration (3)\\ Fixation number (2)\end{tabular} &
  0.616 (0.054) &
  \begin{tabular}[c]{@{}l@{}}Max pupil size (5)\\ Pupil window 1051 (3)\\ Pupil window 351 (3)\\ Mean pupil size (2)\\ Pupil window 1251 (2)\end{tabular} \\ \midrule
\textbf{Revealing / Concealing / Faking} &
   &
   &
   &
   &
   &
   &
   \\ \midrule
Eyelink dataset &
  0.341 &
  0.485 (0.015) &
  \begin{tabular}[c]{@{}l@{}}Saccade amplitude (5)\\ Saccade number (5)\\ Saccade duration (5)\\ Fixation duration (3)\\ Pupil window 1 (2)\end{tabular} &
  0.486 (0.020) &
  \begin{tabular}[c]{@{}l@{}}Saccade amplitude (5) \\ Saccade number (5)\\ Saccade duration (5)\\ Fixation duration (5)\\ Blink duration (4)\end{tabular} &
  0.377 (0.017) &
  \begin{tabular}[c]{@{}l@{}}Min pupil size (5)\\ Mean pupil size (5)\\ Max pupil size (4)\\ Pupil window 1001 (3)\\ Pupil window 2451 (3)\\ Pupil window 801 (2)\\ Pupil window 751 (2)\end{tabular} \\ \midrule
Neon dataset &
  0.361 &
  0.456 (0.064) &
  \begin{tabular}[c]{@{}l@{}}Saccade amplitude (5)\\ Pupil window 1251 (5)\\ Saccade duration (4)\\ Max pupil size (3) \\ Pupil window 951 (2)\end{tabular} &
  0.413 (0.068) &
  \begin{tabular}[c]{@{}l@{}}Saccade amplitude (5)\\ Saccade duration (5)\\ Fixation duration (5)\\ Blink duration (5)\\ Saccade number (3)\end{tabular} &
  0.469 (0.048) &
  \begin{tabular}[c]{@{}l@{}}Max pupil size (5)\\ Pupil window 1251 (4)\\ Pupil window 701 (3)\\ Pupil window 901 (2)\\ Pupil window 951 (2)\\ Pupil window 1001 (2)\\ Pupil window 1051 (2)\end{tabular} \\ \bottomrule
\end{tabular}%
}

\end{table*}

\section{Discussion}

As shown in Table~\ref{tab:results}, the prediction of Concealing and Revealing tasks achieved an accuracy of 74\% on the Eyelink dataset from the full features, and 64\% on the Neon dataset. These results were considerably above the chance level, thereby proving evidence that the eyes can be used as indicators of deceptive intentions. The enhanced accuracy observed in Eyelink dataset could be attributed to an expanded sample size, the use of a chinrest, and a higher sample frequency of the Eyelink than the Neon which should result in a more accurate gaze estimation. However, the Neon dataset exhibited a satisfactory level of accuracy, despite employing a smaller sample size and using gaze data recorded from mobile eye-trackers, which are supposedly associated with diminished data quality. Interestingly, the Eyelink dataset exhibited higher accuracy with eye-movement features, while the Neon dataset demonstrated a stronger association with pupil features. This discrepancy can be interpreted in the context of the disparity in experimental design between the two datasets. The presentation of two cards simultaneously in Eyelink dataset, as opposed to the presentation of a single card in Neon dataset, could have led to an increased number of eye movements in the former. This may have resulted in less reliable pupil measures compared to the latter, where the eyes may have been more focused on the card's area of interest. Nevertheless, these results also serve to reinforce the reliability of mobile eye-trackers in terms of pupil detection.

The prediction of Concealing, Faking, and Revealing tasks achie\-ved an accuracy of 49\% on the Eyelink dataset from the full features and eye features, and 47\% on the Neon dataset with just the pupil features. These results follow the same trend, in that only using eye-movement features to predict in the Neon dataset leads to poor results, while sole use of eye features leads to similar or better results than all features in the Eyelink dataset. The lower accuracy can be attributed to the increased complexity of predicting three classes compared to binary classification. However, previous research already showed that Revealing and Faking tasks in a CIT induce similar eye movements \cite{foucher_unveiling_2024}. While AI enhances differentiation, achieving a clear distinction between these behaviours remains challenging.

Regarding the importance of the eye movement features, among the 60 investigated, saccade number, duration and amplitude consistently constituted the most important features across the two datasets with maximum pupil size being the only pupil feature shared by all feature models and pupil feature models. While it is plausible that other eye parameters may also have contributed to the model's accuracies, these findings serve to strengthen the role of saccades in the prediction of deception in the context of CIT, contradicting previous findings \cite{borza_eye_2018}. Additionally, pupil size had a role in deception prediction with a strong influence on the maximum, minimum and mean pupil size, as well as mean pupil size between 800ms and 1500ms after card presentation. This aligns with prior research \cite{fang_assessing_2021} and underscores the potential of pupil size as a reliable indicator of deception.

These results show the potential of ocular indicators of deception in a study environment. In contrast to our study, lying usually serves a purpose. A higher stakes experimental design (with motivation for successful lies and/or consequences for lying) could increase emotional load and physiological parameters. Furthermore, the robustness of a lie detector system could be improved by coupling eye parameters with other physiological modalities such as brain activity \cite{rosenfeld_p300_2020}.

\paragraph{Ethics and privacy statement.} As mentioned in Section~\ref{sec:methods:participants}, the experiments were approved by the Ethics Committee of Ulm University and all participants provided their written consent. All data were pseudonymised before being processed. While the findings of this study indicate the potential of incorporating gaze and AI into lie detection systems, it is important to acknowledge the limitations of these methods. They should not be relied upon as the sole basis for determining culpability, but rather be regarded as supplementary sources of information that provide guidance.

\section{Conclusion}

To conclude, our findings demonstrated that AI can effectively predict deception from gaze. Specifically, the detection of card concealment achieved accuracies ranging from 64\% in the Neon dataset to 74\% in the Eyelink dataset. The three-classification task proved more complex, with 49\% in the Eyelink dataset and 46\% in the Neon dataset. The feature selection analysis identified saccade number, duration, and amplitude along with maximum pupil size as the most important predictors of deception. These results imply the potential of integrating gaze analysis with AI to enhance lie detection methods and advocate for further research to refine and advance this approach.

\begin{acks}
This project has received funding from the European Union’s Horizon Europe research and innovation funding programme under grant agreement No 101072410. The authors would like to thank Professor Anke Huckauf for participating in elaborating on the initial research idea, and Professor Maria Bielikova for support in the modeling process.
\end{acks}

\bibliographystyle{ACM-Reference-Format}
\bibliography{references}

\begin{figure}[b]
  \includegraphics[scale=0.1]{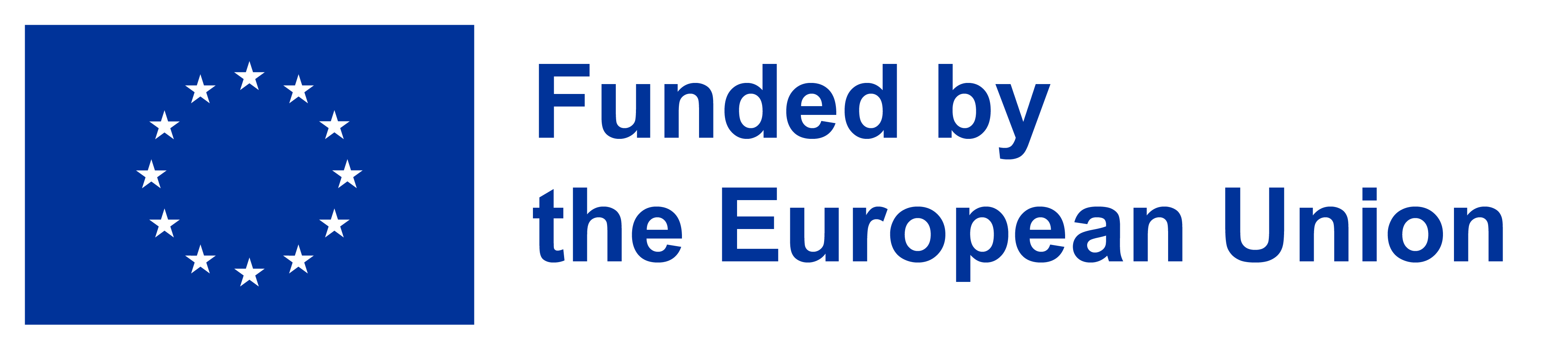}
  \Description{Funded by the European Union}
\end{figure}

\end{document}